\documentclass[a4paper,10pt,twocolumn,twoside,journal]{article}

\usepackage{geometry}
\advance\oddsidemargin by -1in
\advance\evensidemargin by -1in
\advance\headsep by 2.8mm
\def\BibTeX{{\rm B\kern-.05em{\sc i\kern-.025em b}\kern-.08em
    T\kern-.1667em\lower.7ex\hbox{E}\kern-.125emX}}

\usepackage[normalem]{ulem}

\providecommand{\keywords}[1]{\textbf{\textit{Index terms---}} #1}
\usepackage[T1]{fontenc}
\usepackage{graphicx}
\usepackage{braket}
\usepackage{todonotes}
\usepackage{amsfonts, amssymb}
\usepackage{comment}
\usepackage{pgfplots}
\usepackage{multirow}
\usepackage{array}
\usepackage[utf8]{inputenc}
\usepackage{authblk}      
\usepackage{orcidlink}    
\usepackage{hyperref}    

\title{Approximate Quantum State Preparation Through Proximal Policy Optimization}

\author{Marco Mordacci \orcidlink{0009-0001-1955-7197}$^{1,*}$}
\author{Michele Amoretti \orcidlink{0000-0002-6046-1904}$^{1}$}

\affil{\parbox{\textwidth}{\centering
$^1$Quantum Software Laboratory, University of Parma, Parco Area delle Scienze, 181/A, Parma, 43124, Italy.\\
{\upshape(\url{https://www.qslab.unipr.it/})}\\
$^*$Corresponding author: Marco Mordacci, {\upshape\href{mailto:marco.mordacci1@unipr.it}{marco.mordacci1@unipr.it}}
}}

\date{}

\begin{document}
\maketitle        
\begin{abstract}
In this work, a quantum architecture search framework for approximate quantum state preparation (QSP) is proposed. QSP is a challenging task, since the search space grows exponentially with the number of qubits, making the identification of the optimal circuit non-trivial. To address this problem, deep reinforcement learning is employed through an agent based on proximal policy optimization. The objective of the agent is to identify the best possible approximation of the target state while simultaneously minimizing the number of gates used. At each step, the agent appends a new gate to the circuit and recomputes the fidelity between the approximated state and the target states. Various experiments have been performed from 2 to 5 qubits. Both predefined states, such as Bell, GHZ, W, and Dicke states, and completely random states are considered. The proposed framework is able to achieve approximation errors of $10^{-14}$.

\end{abstract}
\section{Introduction}
Quantum computing offers significant potential for solving certain information processing problems that are considered infeasible for classical computers. An important step in many fields is Quantum State Preparation (QSP), which plays a crucial role in several quantum applications, including quantum machine learning~\cite{rath2024quantum}, Hamiltonian simulation~\cite{berry2018improved}, and quantum chemistry~\cite{fomichev2024initial}.

Given a target state $\ket{\psi} = \sum_{i=0}^{2^n-1}a_i\ket{i}$ and the starting state $\ket{0}^{\otimes n}$, QSP aims to find a unitary operator $U$ such that $\ket{\psi} = U\ket{0}^{\otimes n}$, where $\ket{0}^{\otimes n}$ denotes the $n$-qubit initial state. QSP can also be formulated in an approximate setting, where the objective is to construct a state that approximates the target within a given tolerance $\epsilon$, according to a chosen distance metric.

The study of QSP began in 2002, when Grover and Rudolph~\cite{grover2002creating} proposed an algorithm with a depth upper bound of $O(n2^n)$. Subsequently, many works have been published to optimize the circuits~\cite{mottonen2006decompositions,plesch2011quantum}. Furthermore, research on QSP based on ancillae started~\cite{zhang2021generic,sun2023asymptotically}.

In the approximate case, methods based on variational quantum circuits (VQCs)~\cite{benedetti2019parameterized} were proposed~\cite{hai2023variational}. 
In 2025, Rofougaran et al.~\cite{rofougaran2025encoding} provided an algorithm to encode proteins as quantum states, and Belli et al.~\cite{belli2025} proposed a method based on the standard recursive block basis~\cite{belli2024,Belli_2026} to prepare an arbitrary quantum state.  

Another important research direction is Quantum Architecture Search (QAS)~\cite{Martyniuk_2024}, which encompasses a broad class of techniques aimed at automating the design of optimal parameterized quantum circuits (PQCs). The manual design of PQCs is a non-trivial task, since the principles that determine why a circuit architecture performs well for a given application are still not fully understood and remain an active area of research. For instance, in quantum machine learning, predefined architectures are used~\cite{Leone_2024}; however, understanding why certain circuit structures outperform others and the relationship between circuit topology, expressibility, trainability, and performance is still an open problem ~\cite{IlarioCorrer_2025,mordacci2025}.

QAS leverages various optimization strategies, such as evolutionary algorithms, reinforcement learning, differentiable methods, and Bayesian optimization, to automatically discover efficient circuit architectures.

Several studies have applied QAS techniques to QSP. Reinforcement learning approaches were used for Bell and GHZ state preparation~\cite{kuo2021quantumarchitecturesearchdeep,Zhu2023QuantumAS}, while genetic algorithms were explored for GHZ, Gaussian, and W states in~\cite{10821312,Creevey2023GASP}. More recently, diffusion and generative models have also been investigated for quantum circuit synthesis and state preparation tasks~\cite{11250254,rapp2025reinforcement,kolle2025evaluating}.

This paper provides the following new contributions: 1) a novel QSP algorithm based on proximal policy approximation; 2) the assessment of the algorithm's performance on both predefined states and random states.

The remainder of the paper is organized as follows. In Section~\ref{lab:th_frame}, the proposed architecture is presented in detail, including its main components and the underlying theoretical background. In particular, the section introduces the fundamentals of reinforcement learning, the Actor-Critic paradigm, proximal policy optimization, and, finally, the proposed architecture for approximate QSP. In Section~\ref{lab:res}, the results from 2 to 5 qubits are presented. Both predefined states, such as Bell, GHZ, W, and Dicke states, and random states are considered. Finally, Section~\ref{lab:conclusion} concludes the paper with a summary of the main results and a discussion of future work.

\section{Theoretical Framework}
\label{lab:th_frame}
\subsection{Reinforcement Learning}
Reinforcement Learning (RL)~\cite{sutton1998reinforcement} is a machine learning paradigm in which an agent learns to make decisions by interacting with an environment. During this interaction, the agent operates over a sequence of discrete time steps. At each step $t$, the environment provides the agent with a state or observation $s_t$. based on this information, the agent selects an action $a_t$ from a set of possible actions $A$ according to a policy $\pi$. The policy $\pi$ defines how actions are chosen from states, which means that for a given state $s_t$, it outputs a probability distribution $\pi(a_t|s_t)$ over the possible actions. Once the action $a_t$ is executed, the environment responds by providing the next state $s_{t+1}$ with a scalar value $r_t$, called reward, which evaluates the quality of the chosen actions. The learning process is typically divided into two approaches: value-based methods, which learn to estimate the expected cumulative reward (the value) of being in a state or taking an action, and policy-based methods, which directly optimize the policy to maximize the total reward. This process continues until either a terminal state or a stopping condition, such as a maximum number of steps, is reached. An episode corresponds to a sequence of actions selected by the agent, observed states, and reward until a termination condition is reached.

\subsection{Actor-Critic and Proximal Policy Optimization}
Actor-Critic~\cite{konda1999actor} methods represent a hybrid architecture that combines the advantages of both policy-based and value-based reinforcement learning. This framework consists of two separate components: the Actor, which is responsible for selecting actions by learning a policy $\pi(a_t, s_t)$, and the Critic, which evaluates the quality of those actions by estimating a value function or expected future rewards. During training, the critic provides feedback to the actor. This feedback is then used to update both components: the critic is updated to be more precise in its feedback, while the actor updates its policy parameters in a direction that favors actions with positive feedback. By utilizing a critic to reduce the variance of the policy gradient, these methods often achieve more stable and efficient learning compared to pure policy gradient approaches.

Proximal Policy Optimization (PPO)~\cite{schulman2017proximal} is a policy-gradient method designed to improve training stability and sample efficiency. To achieve this, PPO constrains the policy updates so that the new policy does not deviate excessively from the previous one. This is obtained through the clipped surrogate objective function:
\begin{equation}
L^{\mathrm{CLIP}}(\theta) =
\mathbb{E}_t \left[
\min \left(
r_t(\theta)\hat{A}_t,\;
\mathrm{clip}\left(r_t(\theta), 1-\epsilon, 1+\epsilon\right)\hat{A}_t
\right)
\right]
\end{equation}
with
\begin{equation}
r_t(\theta)=
\frac{\pi_\theta(a_t|s_t)}
{\pi_{\theta_{\mathrm{old}}}(a_t|s_t)}
\end{equation}
where $r_t(\theta)$ measures how much the probability of taking a given action changes after the update, $\hat{A}_t$ is the estimated advantage, indicating how good an action $a_t$ is, and $\epsilon$ is a hyperparameter that controls the maximum variation of the policy. The clip restricts $r_t(\theta)$ to the interval $\left[1-\epsilon,1+\epsilon\right]$, avoiding  destructive changes that could destabilize training. As a result, PPO encourages actions with positive advantages while discouraging excessively large parameter changes, leading to more robust and stable learning performance.

\subsection{Proposed Architecture}
The proposed framework is based on a PPO agent implemented within an Actor-Critic architecture. The objective of the agent is to sequentially construct a PQC capable of approximating a target quantum state while minimizing the number of gates employed through interactions with a quantum environment.

The actor and critic are implemented using neural networks. Both consist of three fully connected layers: $Input \rightarrow 256 \rightarrow 256 \rightarrow output$. The actor network has a final softmax layer that produces the action probability distribution, while the critic network outputs a scalar state-value estimate.

The agent progressively builds a quantum circuit gate-by-gate. In each interaction with the environment, the agent selects a quantum operation from a predefined discrete action space composed of single-qubit rotation gates ($R_x(\theta)$, $R_y(\theta)$, $R_z(\theta)$) and CNOT gates.
For an $n$-qubit system, the action space contains all admissible gate-wire combinations. Whenever the PPO agent adds a new gate to the circuit, the environment begins a small training of the rotation angles. The optimization minimizes a cost function based on the fidelity between the generated and the target state. The following loss function is used:
\begin{equation}
    L = 1 - F(\psi, \phi)
\end{equation}
where $F(\psi, \phi) = |\braket{\psi|\phi}|^2$ with $\psi$ as the target state and $\phi$ as the current state. Early stopping is implemented to interrupt optimization when the improvement becomes negligible, reducing computational overhead.

The observation returned to the agent is a tuple composed of the current quantum state reached, the number of gates used, and the difference between the target and the current quantum state.

The framework adopts a logarithmic reward function based on the fidelity in order to better guide the agent during high-precision optimization. When the generated quantum state becomes very close to the target state, the fidelity error can reach extremely small values, such as $10^{-10}$. In this regime, a linear reward provides only small variations between successive improvements, making it difficult for the agent to distinguish meaningful progress. By using a logarithmic reward, small reductions in the error are amplified, allowing the agent to remain sensitive to incremental improvements even at very high fidelities and facilitating convergence toward precise quantum state approximations. The reward is computed as:
\begin{equation}
    r_t = \lambda\left[ -log(D_t + \epsilon) + log(D_{t-1} + \epsilon)\right] - p
\end{equation}
where $D_t$ is the current distance, $D_{t-1}$ is the previous distance, $\lambda$ is a scaling coefficient, $p$ is a penalty factor used to discourage the generation of unnecessary deep circuits, and $\epsilon$ is a small constant set to $10^{-16}$.

\section{Results}
\label{lab:res}
The proposed architecture is implemented using the PennyLane library~\cite{bergholm2018pennylane}. The tests are performed using the simulator provided by the library, executed on a Linux machine equipped with an AMD EPYC 7282 CPU and 256 GB of RAM, supplied by the HPC facility of the University of Parma. Moreover, the tests are executed using 16 parallel environments for states with more than two qubits, while only a single environment is employed for the two-qubit experiments. The agent is updated every $128$ steps.

The penalty factor $p$ and the multiplier $\lambda$ are set to $10$. The threshold that the agent must reach in order to receive the final reward of $50$ is set to $10^{-14}$. The action space is composed of the three possible single-qubit rotations ($R_x$, $R_y$, and $R_z$) and the CNOT gate as entangling gates. Furthermore, whenever a new gate is appended to the circuit, a short training is performed to determine the optimal $\theta$ parameters of the circuit. 
For two-qubit quantum states, the maximum number of gates is limited to 20; for three qubits, the limit is set to 30 gates; with four qubits, the limit is increased to 70 gates for random states, while for W and Dicke states it is restricted to 40 gates.

The Adam optimizer is used to train both the agent and the internal quantum circuit, with learning rates $10^{-4}$ and $0.05$, respectively.

\begin{figure*}[htbp]
    \centering

    \begin{minipage}{0.7\linewidth}
        \centering
        \makebox[0pt][l]{\textbf{a)}}\\
        \includegraphics[width=\linewidth]{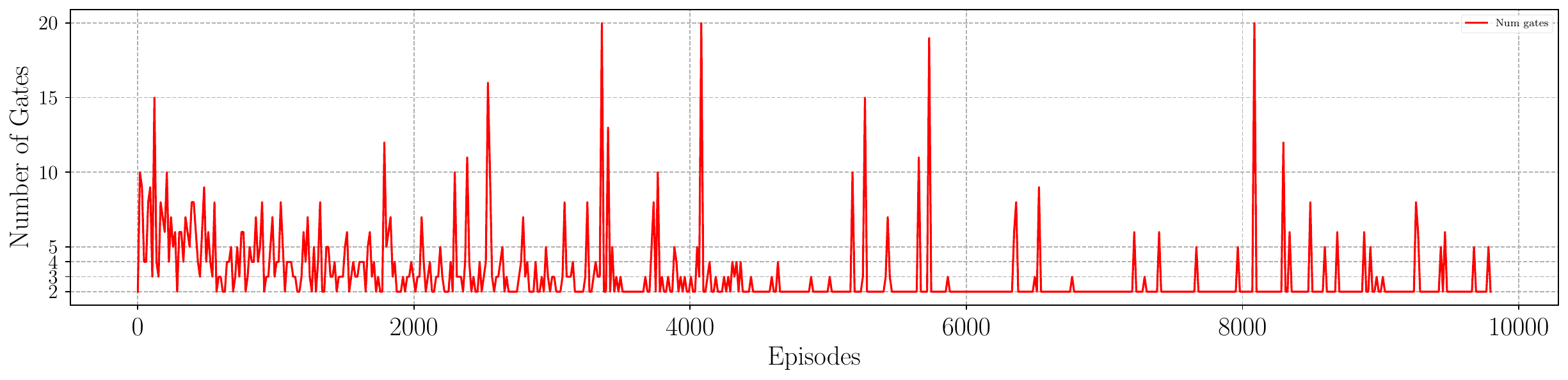}
        
        \vspace{0.3cm}
        
        \includegraphics[width=\linewidth]{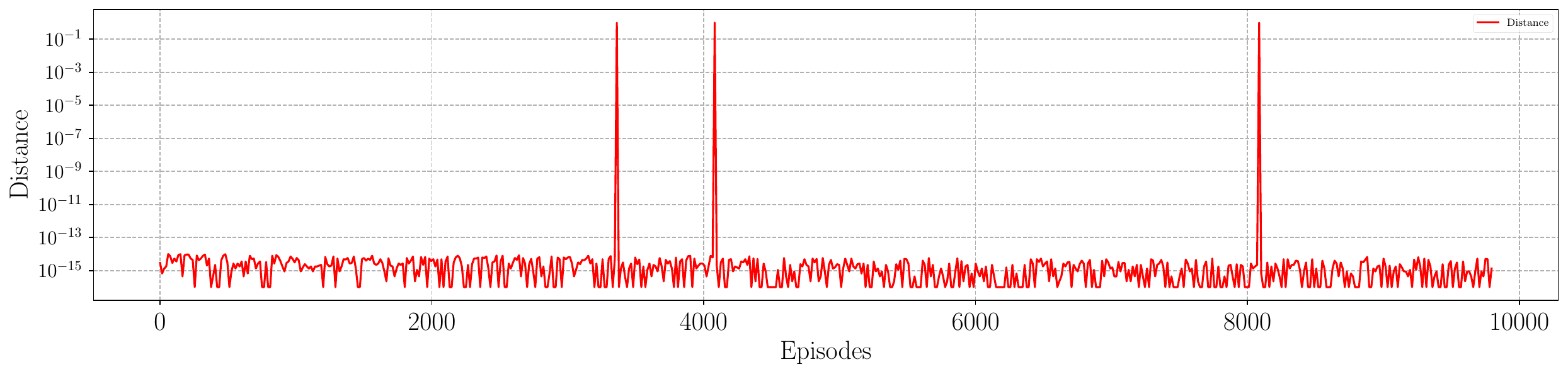}
    \end{minipage}
    \hfill
    \begin{minipage}{0.28\linewidth}
        \centering
        \makebox[0pt][l]{\textbf{b)}}\\
        \includegraphics[width=\linewidth]{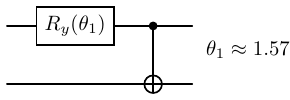}
    \end{minipage}

    \caption{(a) Number of gates and trace distance achieved during training for the Bell state. (b) Circuit learned by the agent to approximate the Bell state.}
    \label{fig:bell_full}
\end{figure*}

For two-qubit quantum states, both predefined states and random states are analyzed. The approximation error and the number of gates achieved throughout the training episodes are reported and discussed. In particular, Figure~\ref{fig:bell_full}.a presents the evolution of the approximation error and the number of gates obtained during the training process for the Bell state; the obtained circuit is sampled every 15 episodes. The distance plot shows how the agent is able to find the $10^{-14}$ approximation from the start, while to find the optimal solution, it needs around $5000$ episodes. Moreover, by analyzing the generated quantum circuit shown in Figure~\ref{fig:bell_full}.b, it can be observed that the exact solution $H(q_0)\otimes CNOT(q_0,q_1)$ is not found since the Hadamard gate is not included in the action space. Instead, its effect is approximated through the gate $R_y(1.57)$. In some cases, the agent does not converge to the optimal solution. This happens because it identifies $CNOT(q_0, q_1)$ and $R_y$ as the best gates; however, due to insufficient penalization, it may overuse them in certain runs. For example, it could find the solution $R_y(q_0) \otimes R_y(q_0) \otimes CNOT(q_0, q_1)$, which, with an intermediate step, can be compressed to $R_y \otimes CNOT$ by merging the two $R_y$.

\begin{figure*}[htbp]
    \centering
    \begin{minipage}{0.7\linewidth}
    \makebox[0pt][l]{\textbf{a)}}\\
    \includegraphics[width=\linewidth]{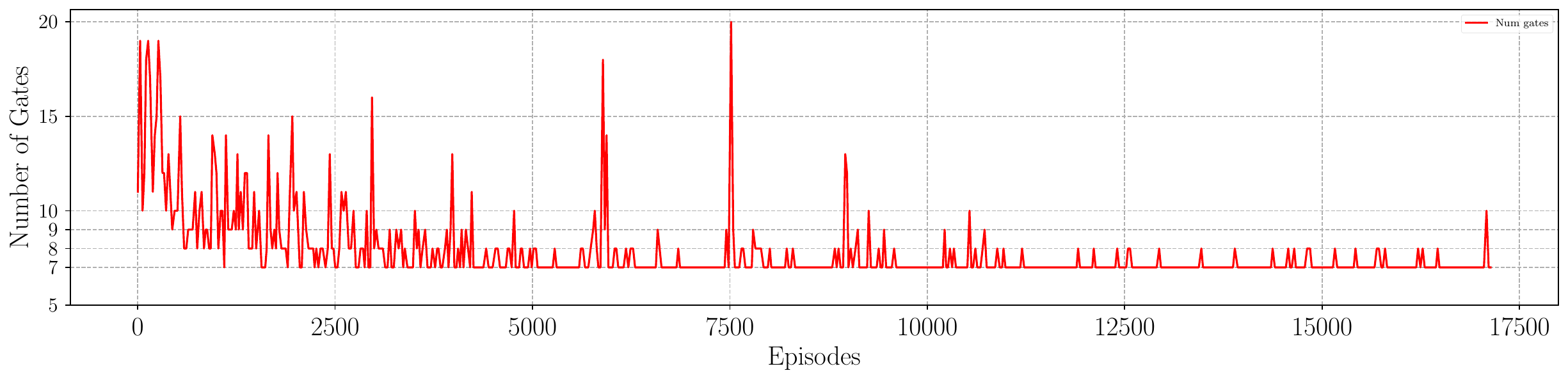}
    \includegraphics[width=\linewidth]{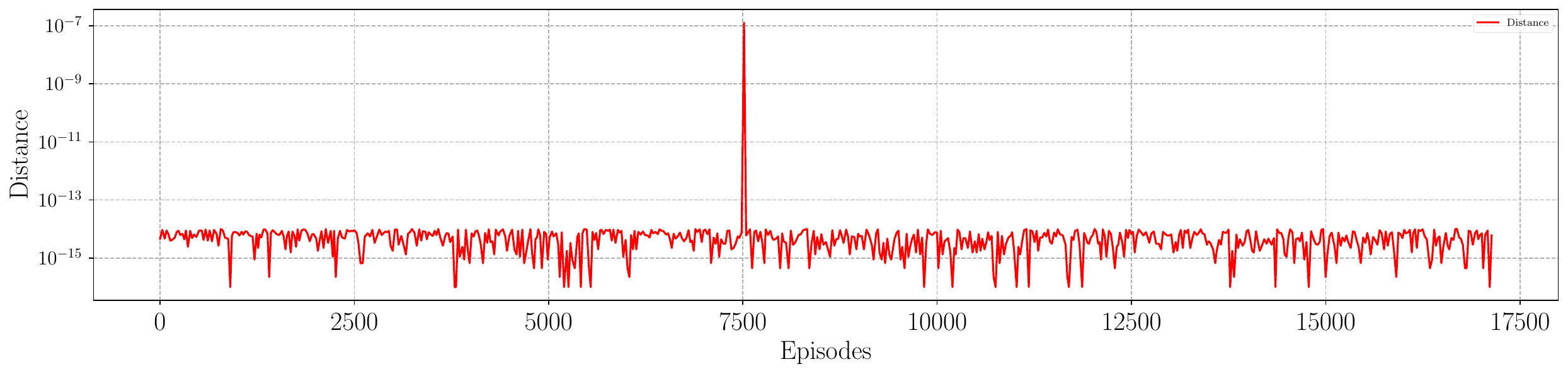}
    \end{minipage}
    \begin{minipage}{0.28\linewidth}
    \makebox[0pt][l]{\textbf{b)}}\\
    \includegraphics[width=\linewidth]{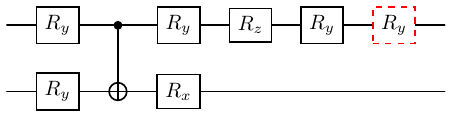}
    \end{minipage}
    \caption{a) Number of gates and trace distance achieved during the training for a random 2-qubit state. b) Circuit achieved by the agent to approximate a 2-qubit random state.}
    \label{fig:r2Res}
\end{figure*}

In Figure~\ref{fig:r2Res}.a, an example of the execution on a two-qubit random state is presented. In this case as well, the agent is capable of finding a valid solution in almost every episode, while the optimal solution composed of $7$ gates starts to be consistently identified after around $4000$ episodes. This result is consistent with the findings reported  in~\cite{plesch2011quantum}, from which it can be derived that $7$ gates represent the minimum number of gates required to prepare an arbitrary two-qubit quantum state. Furthermore, Figure~\ref{fig:r2Res}.b shows an example of a generated circuit. In most cases, the circuit produced is the optimal one, without the dashed $R_y$. However, occasionally, the agent produces the circuit with $8$ gates because the early stopping procedure is overly restrictive and the training is stopped too early. Nevertheless, this circuit can still be simplified by merging the two $R_y$ gates. Even though the optimal solution is not found automatically by the agent at every run, it is better to maintain early stopping since the training time for one update of the agent is reduced from around $230$ to $60$ seconds.

\begin{figure*}[h]
    \centering

    \begin{minipage}{0.7\linewidth}
        \centering
        \makebox[0pt][l]{\textbf{a)}}\\
        \includegraphics[width=\linewidth]{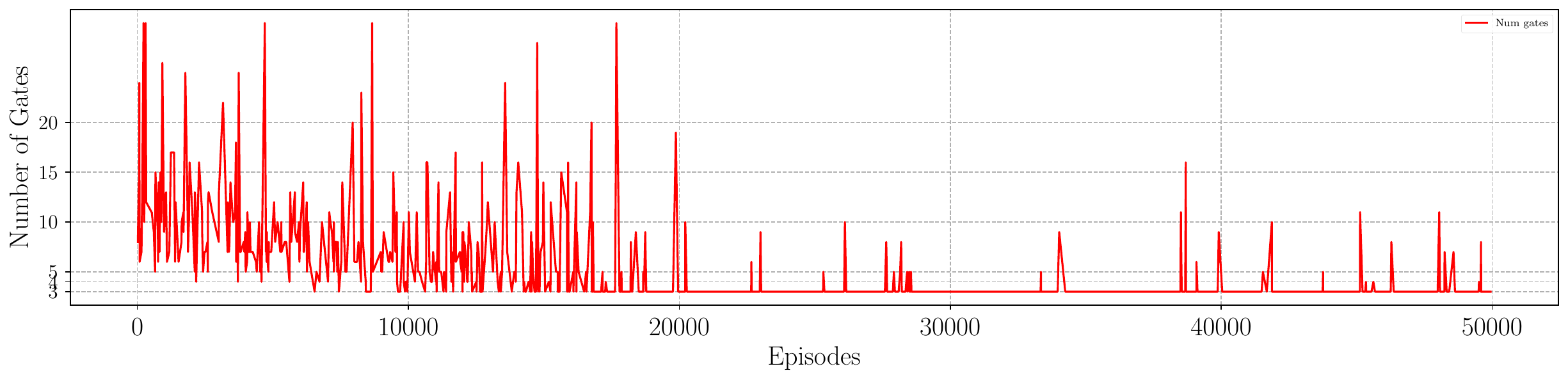}
        \includegraphics[width=\linewidth]{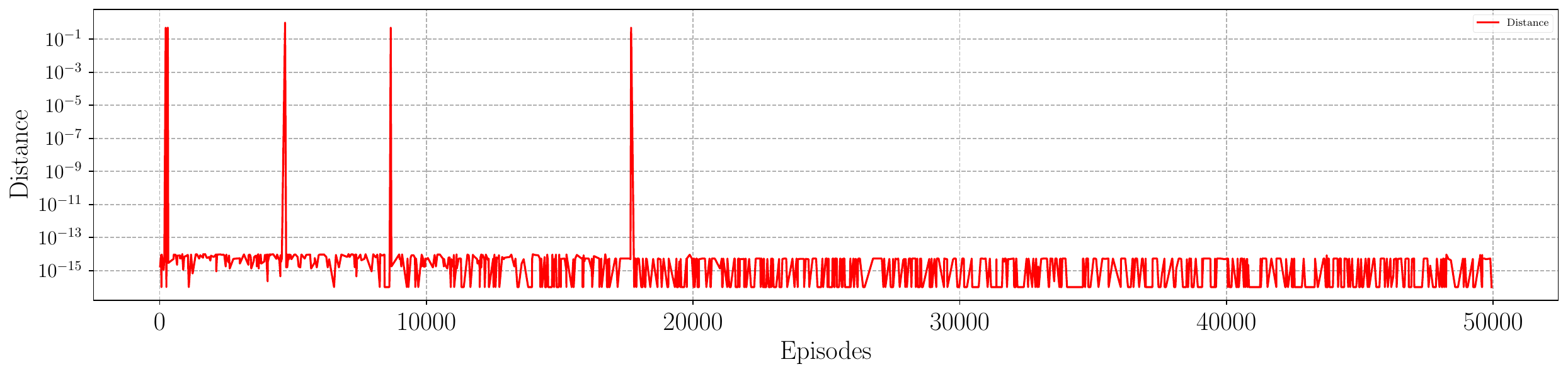}
    \end{minipage}
    \hfill
    \begin{minipage}{0.25\linewidth}
        \centering
        \makebox[0pt][l]{\textbf{c)}}\\
        \includegraphics[width=0.8\linewidth]{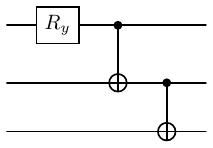}
    \end{minipage}

    \vspace{0.5cm}

    \begin{minipage}{0.7\linewidth}
        \centering
        \makebox[0pt][l]{\textbf{b)}}\\
        \includegraphics[width=\linewidth]{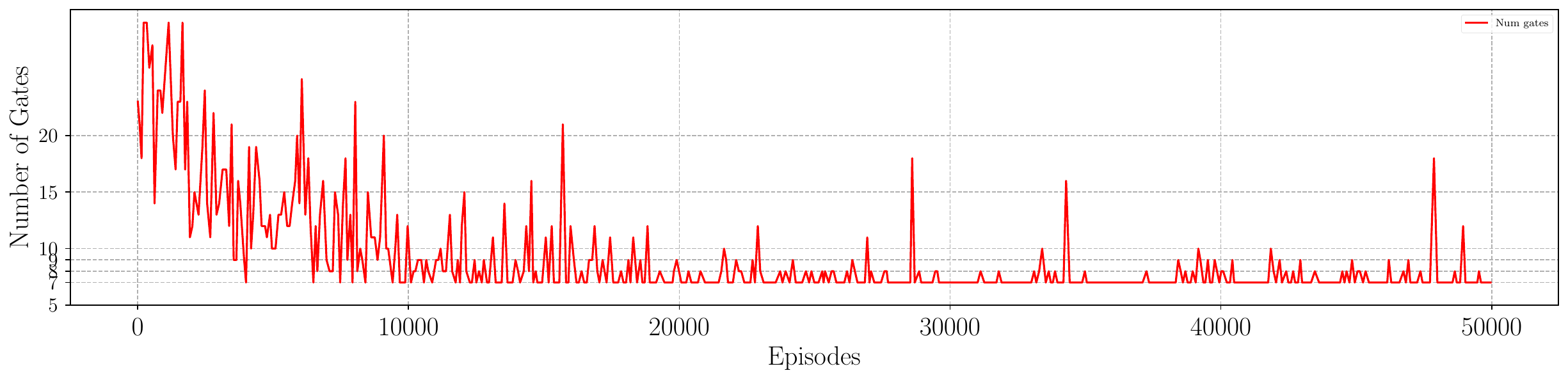}
        \includegraphics[width=\linewidth]{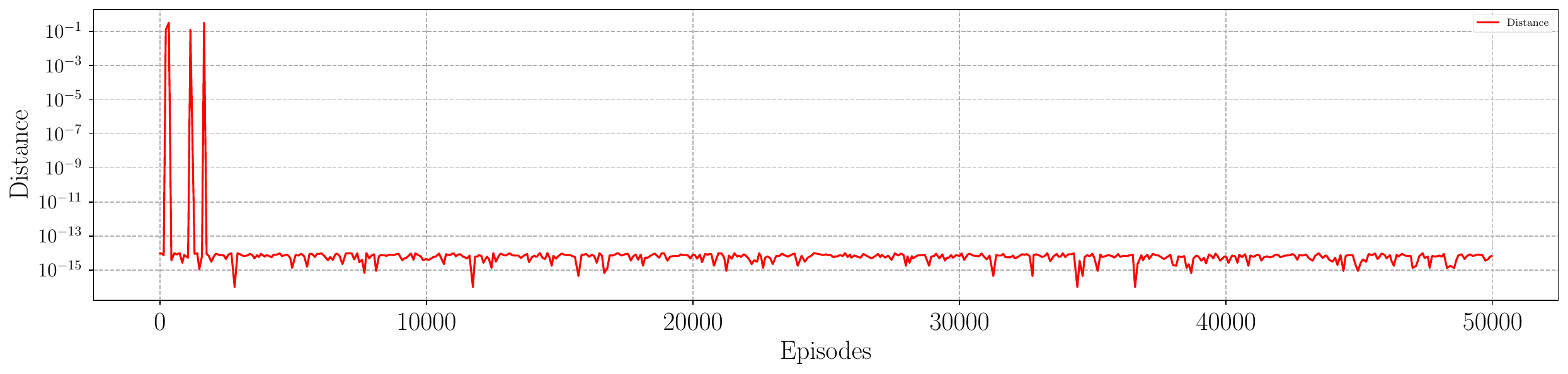}
    \end{minipage}
    \hfill
    \begin{minipage}{0.25\linewidth}
        \centering
        \makebox[0pt][l]{\textbf{d)}}\\
        \includegraphics[width=0.8\linewidth]{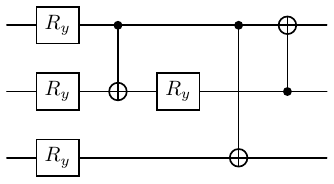}
    \end{minipage}

    \caption{Number of gates and trace distance achieved during the training for a) GHZ and b) W states. Furthermore, the circuits for c) GHZ and d) W states are presented.}
    \label{fig:GHZRes}
\end{figure*}

With three qubits, GHZ and W are tested as predefined states. In Figure~\ref{fig:GHZRes}, the results for both states are presented. The plots show that the results achieved with 2 qubits also hold for them. Indeed, the GHZ state is approximated with the best possible circuit, while the W state circuit is consistent with the literature. Furthermore, the produced quantum circuits are presented.

\begin{figure*}[htbp]
    \centering
    \makebox[0pt][l]{\textbf{a)}}\\
    \includegraphics[width=0.8\linewidth]{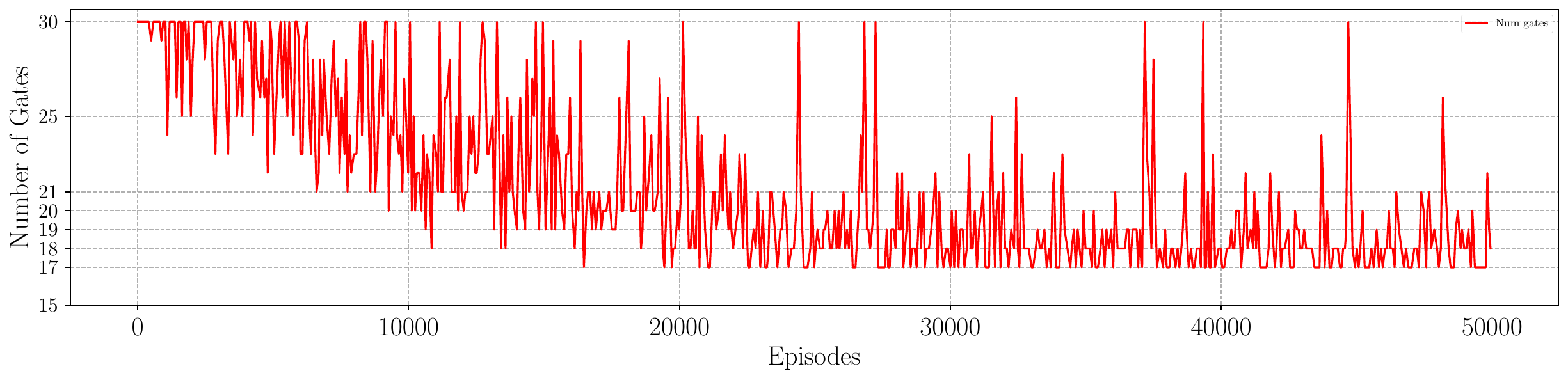}
    \includegraphics[width=0.8\linewidth]{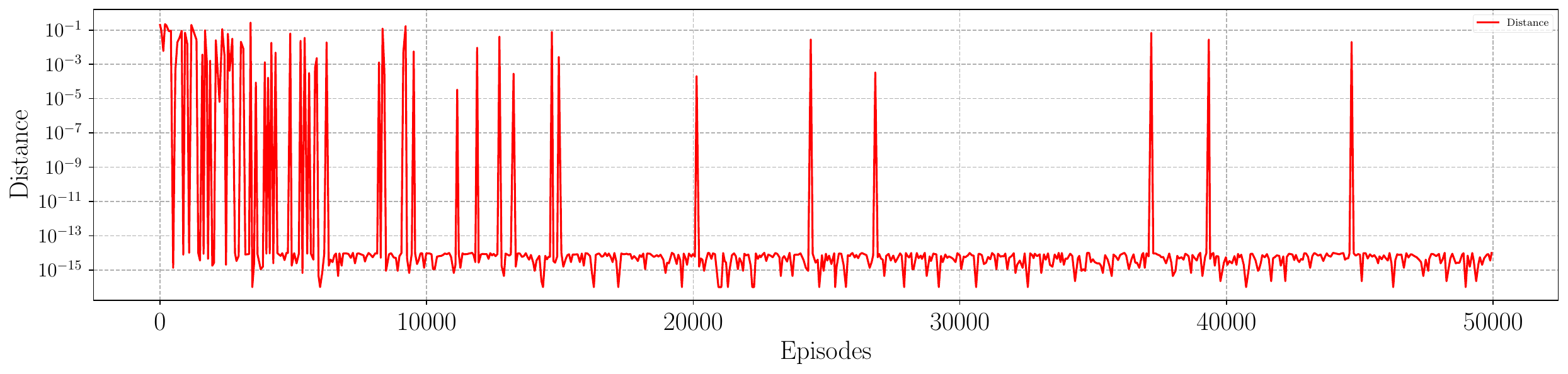}\\
    \makebox[0pt][l]{\textbf{b)}}\\
    \includegraphics[width=0.9\linewidth]{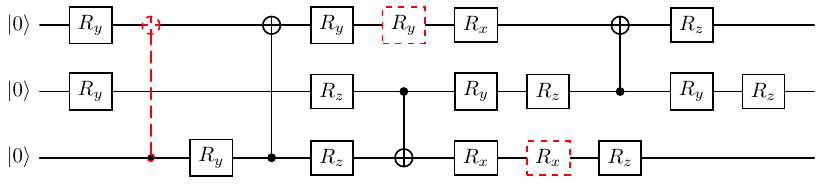}
    \caption{a) Number of gates and trace distance achieved during the training for a 3-qubit random state. b) Circuit achieved by the agent to approximate a random 3-qubit state.}
    \label{fig:R3Res}
\end{figure*}

\begin{figure*}[htbp]
    \centering
    \includegraphics[width=0.8\linewidth]{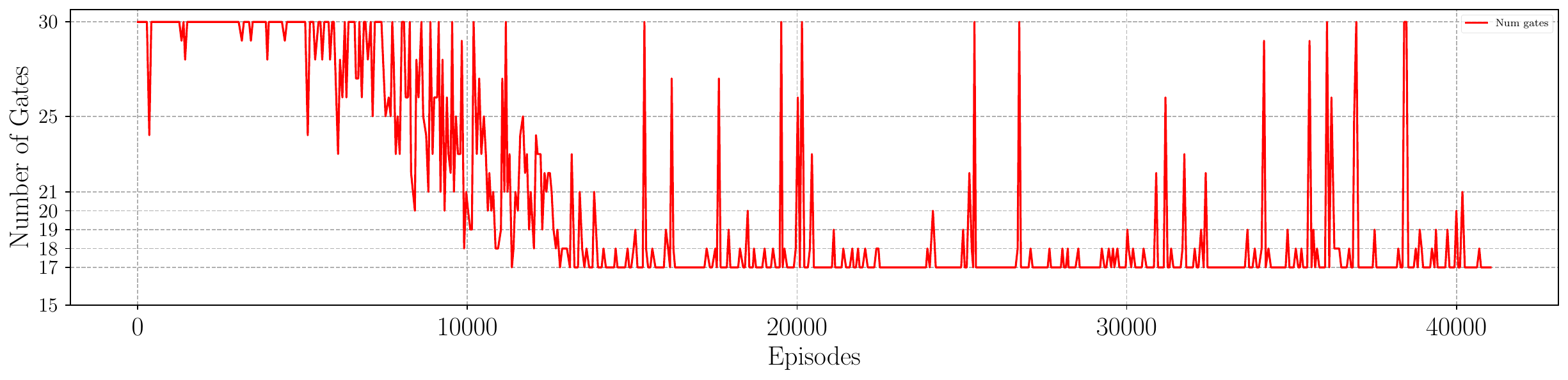}
    \caption{Number of gates achieved during the training for a 3-qubit random state with $\lambda=1$ and $p=5\times\#AppendedGates$.}
    \label{fig:R3_2Res}
\end{figure*}

Subsequently, various three-qubit random states are evaluated. Figure~\ref{fig:R3Res}.a presents an example of execution. The reported trends show that, with respect to the trace distance, the results achieved with the previous tests still hold, as the agent is consistently capable of achieving very accurate approximations of the target state. However, the number of gates begins to exhibit larger fluctuations, and the minimum circuit configuration is not identified consistently. In particular, it was shown in~\cite{vznidarivc2008optimal,giraud2009quantum,plesch2011quantum} that the minimum number of gates required to prepare an arbitrary three-qubit random state is $17$, consisting of $3$ CNOT and $14$ parameterized rotations. The proposed algorithm, instead, tends to oscillate between approximately $17$ and $20$ gates on average. Nevertheless, in most cases, even sub-optimal solutions can be simplified to the optimal one by merging or removing redundant gates, such as in Figure~\ref{fig:R3Res}.b. Furthermore, to improve stability around the optimal solution, an additional configuration with $\lambda=1$ and $p=5\times\#AppendedGates$, where the penalty increases proportionally to the number of currently appended gates, is considered. As shown in Figure~\ref{fig:R3_2Res}, the agent is able to find the $17$-gate solution faster. However, in some cases, the penalty becomes excessively restrictive, and the agent never finds the solution. This behavior is observed in 1 out of 10 runs.

\begin{figure*}[htbp]
    \centering
    \makebox[0pt][l]{\textbf{a)}}\\
    \includegraphics[width=0.8\linewidth]{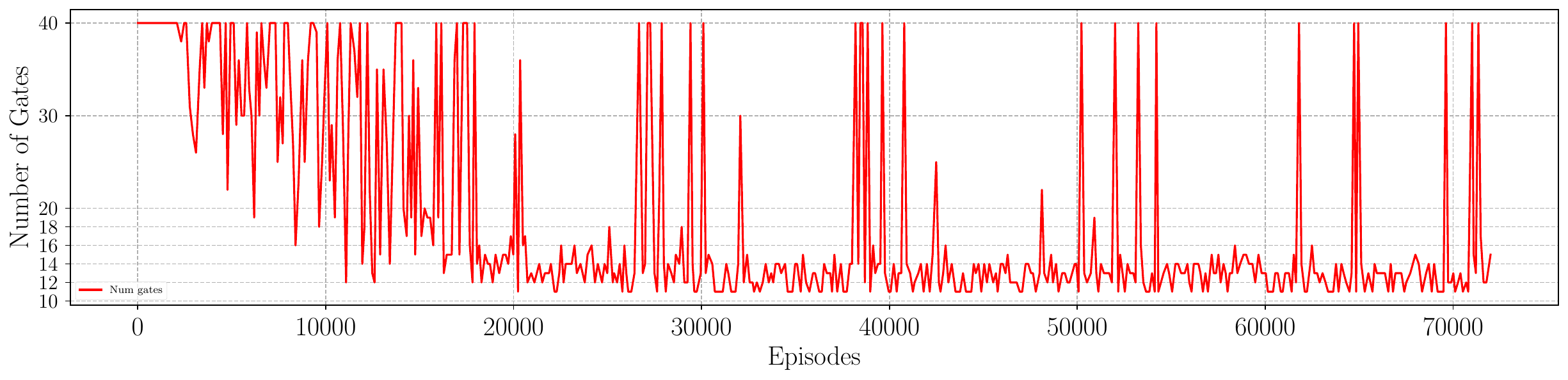}
    \includegraphics[width=0.8\linewidth]{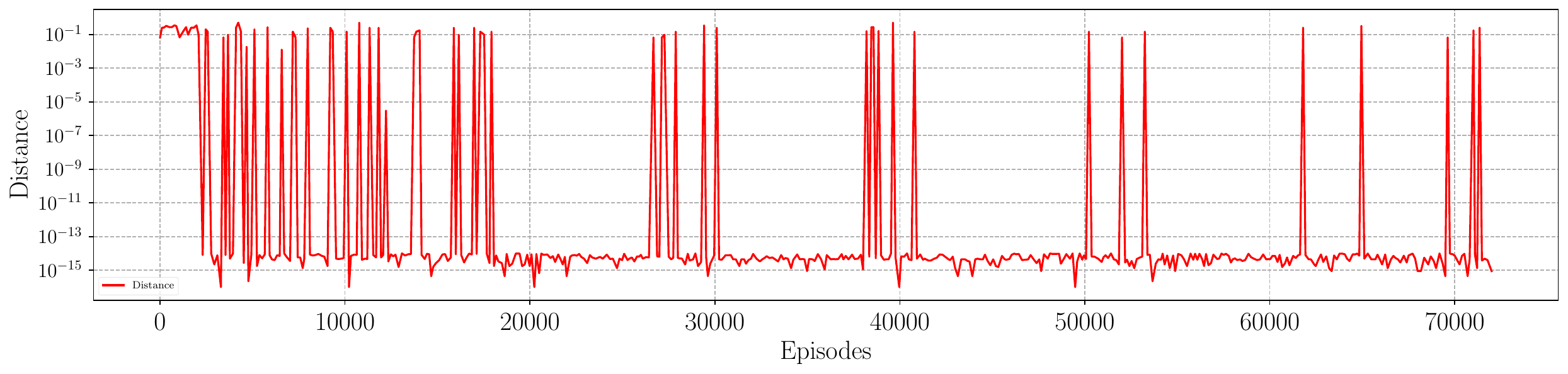}\\
    \makebox[0pt][l]{\textbf{b)}}\\
    \includegraphics[width=0.8\linewidth]{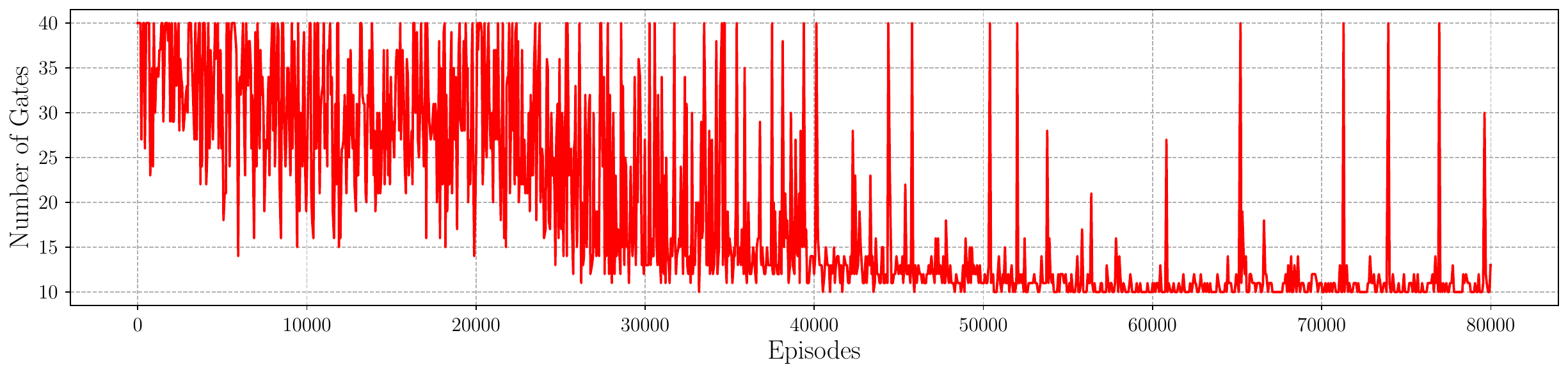}
    \includegraphics[width=0.8\linewidth]{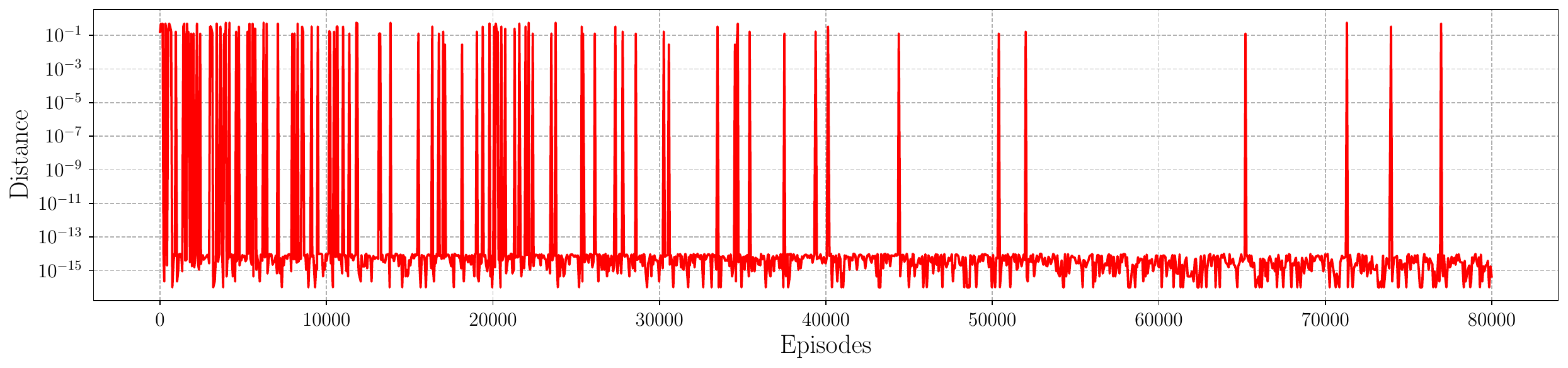}
    \caption{Number of gates and trace distance achieved during the training for W and Dicke$_2$ states.}
    \label{fig:W_dicke_res}
\end{figure*}

\begin{figure*}[htbp]
    \centering
    \includegraphics[width=0.8\linewidth]{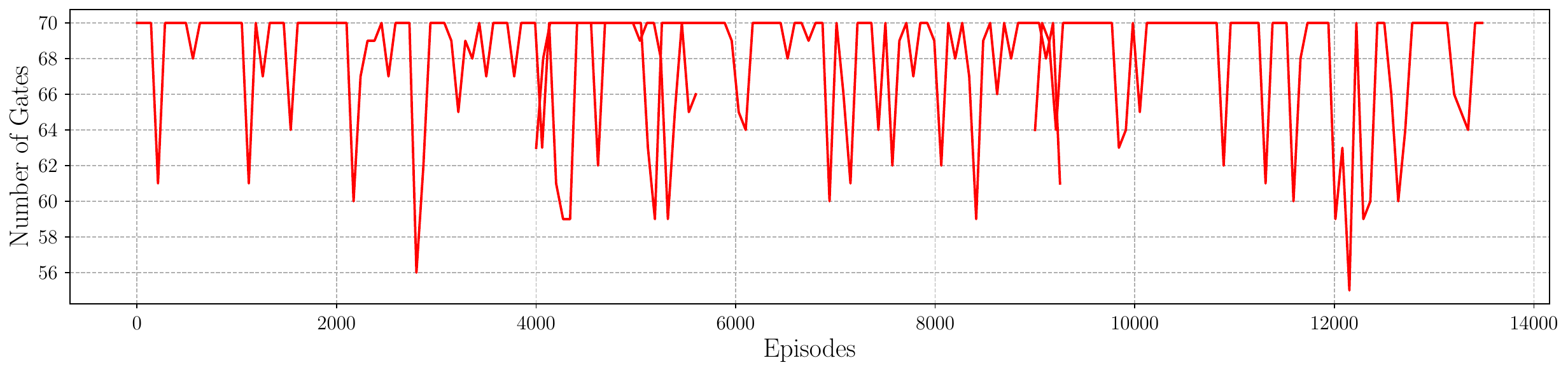}
    \includegraphics[width=0.8\linewidth]{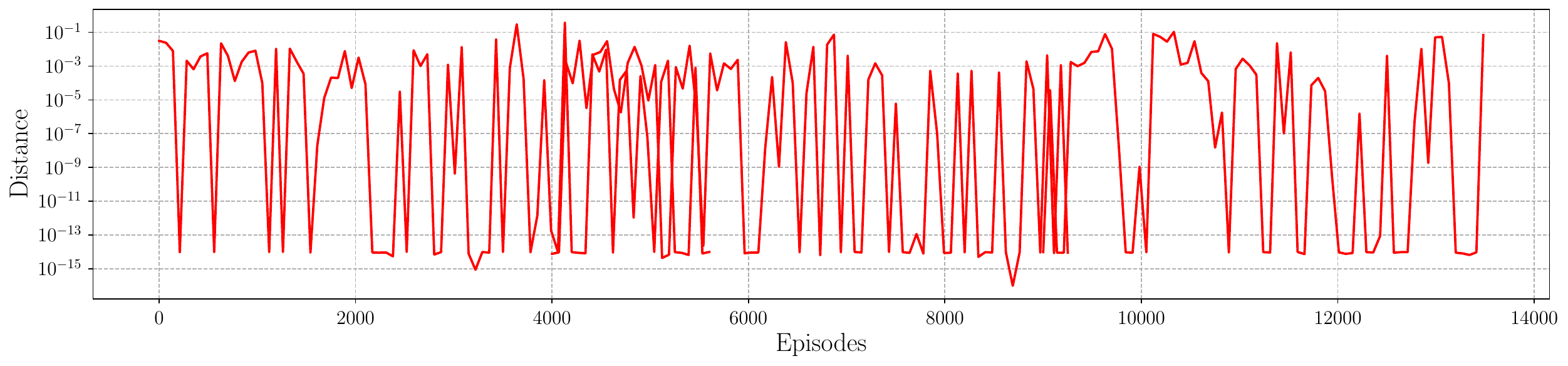}
    \caption{Number of gates and trace distance achieved during the training for a 4-qubit random state.}
    \label{fig:R4Res}
\end{figure*}

Finally, four-qubit states are tested. In Figure~\ref{fig:W_dicke_res}, the results for the W and Dicke$_2$ states are presented. In this case, since the standard configuration does not lead to stable performance, a success buffer is used to improve the agent. This buffer contains the $10$ best circuits found by the agent. Every $5\times \#steps = 5 \times 128$, one solution is randomly sampled and used to perform a short training of the agent. The agent is capable of finding solutions composed of $11\sim16$ gates for the W state, while circuits with $10\sim12$ gates are obtained for the Dicke$_2$ state. Furthermore, Figure~\ref{fig:R4Res} reports the performance for random states. In this case, the agent, after $10000$ episodes, is still not able to consistently find the state precisely but is starting to learn.

In Table~\ref{tab:results_qubits}, the results are summarized. The \textbf{\#gates} column indicates the results without any simplification in the circuit, while \textbf{Optimized \#gates} indicates the results when the circuits are compressed after training. Furthermore, the results on $5$ qubits are presented. The agent can consistently identify a solution since the gate limit is large; however, the model is not fully trained in this scenario. The results on $5$ qubits are included to demonstrate that the framework can be extended to a larger number of qubits.

\begin{table}[htbp]
    \centering
    \caption{The results achieved from 2 to 5 qubits}
    
    \resizebox{0.8\linewidth}{!}{%
    \begin{tabular}{|c|c|c|c|}
        \hline
        \textbf{N} & \textbf{State} & \textbf{\#gates} & \textbf{Optimized \#gates} \\
        \hline
        \multirow{2}{*}{2} 
            & Bell     & $2$ & - \\
            & R2       & $7$ & -\\
        \hline
        \multirow{3}{*}{3} 
            & W         &  $7$ & -\\
            & GHZ       &  $3$ & -\\
            & R3        &  $17\sim20$ & $17\sim18$\\
        \hline
        \multirow{3}{*}{4} 
            & W4        & $11\sim16$& $10\sim15$ \\
            & Dicke2    & $10\sim12$& - \\
            & R4        & $47\sim70($mean$\approx63)$ &  $45\sim60($mean$\approx53)$\\
        \hline
        5 
            & R5        & $140\sim190$ &- \\
        \hline
    \end{tabular}%
    }

    \label{tab:results_qubits}
\end{table}

\section{Conclusion}

\label{lab:conclusion}
In this work, a novel algorithm for QSP is proposed. The approach is based on reinforcement learning and, in particular, employs an agent built upon the PPO algorithm and the Actor-Critic architecture.

The objective of the agent is to iteratively construct the parameterized quantum circuit to produce the target state given as input. This is performed by appending a new gate to the circuit at each step. The action space consists of the three single-qubit rotations ($R_x$, $R_y$, and $R_z$), and the CNOT gate. When a new action is performed, short training on the PQC is conducted to find the optimal configuration.

The framework is evaluated on various predefined and random states ranging from 2 to 5 qubits. The trace distance threshold that must be reached by the agent is $10^{-14}$. 
For each configuration, both the quality of the obtained approximation and the complexity of the generated quantum circuit are analyzed. In particular, the final approximation error and the number of gates selected by the agent are discussed.
The results show that the agent is able to find optimal solutions for a large number of two- and three-qubit states, considering both predefined and random states. However, the performance deteriorates as the number of qubits increases due to the exponential growth of the search space and the corresponding increase in circuit complexity.

In the future, the proposed framework will be modified to remove the internal training of the PQC to achieve more stable results in the identification of the best circuits and to have faster training of the agent. Furthermore, the approach will be tested with a larger number of qubits and compared with existing QSP methods in terms of quantum circuit depth.

\section*{Acknowledgement}
This research benefits from the HPC (High Performance Computing) facility of the University of Parma, Italy.
%
%
%
%
\bibliographystyle{plain}
\bibliography{bibliography}

\end{document}